\documentclass[prd,twocolumn,notitlepage,showpacs,preprintnumbers,amsmath,amssymb,nofootinbib,APS,10pt,superscriptaddress]{revtex4-1}

\usepackage{dcolumn}
\usepackage{bm}
\usepackage[applemac]{inputenc}
\usepackage[spanish,english]{babel}
\usepackage{amsmath,amssymb,amsfonts,latexsym,cancel}
\usepackage{graphicx}
\usepackage{color}
\usepackage{soul}
\usepackage{ulem}
\usepackage{hyperref}

\allowdisplaybreaks

\newcommand{\be}{\begin{equation}}
\newcommand{\ee}{\end{equation}}
\newcommand{\bea}{\begin{eqnarray}}
\newcommand{\eea}{\end{eqnarray}}

\begin{document}
\title{{\bf
Running couplings from adiabatic regularization}}

\author{Antonio Ferreiro}\email{antonio.ferreiro@ific.uv.es}
%\affiliation{Departamento de Fisica Teorica and IFIC, Centro Mixto Universidad de Valencia-CSIC. Facultad de Fisica, Universidad de Valencia, Burjassot-46100, Valencia, Spain.}
%\affiliation{Department of Physics, University of Basel, Klingelbergstr. 82, CH-4056 Basel, Switzerland}

\author{Jose Navarro-Salas}\email{jnavarro@ific.uv.es}
\affiliation{Departamento de F\'isica Te\'orica and IFIC, Centro Mixto Universidad de Valencia-CSIC. Facultad de F\'isica, Universidad de Valencia, Burjassot-46100, Valencia, Spain.}

\begin{abstract}

We extend the adiabatic regularization method by introducing an arbitrary mass scale $\mu$ in the construction of the subtraction terms. This allows us to obtain, in a very robust way, the running of the coupling constants by demanding $\mu$-invariance of the effective semiclassical (Maxwell-Einstein) equations. In particular,  we get the  running of the electric charge of perturbative quantum electrodynamics.  Furthermore, the method  brings about a renormalization of the  cosmological constant and the Newtonian gravitational constant.  The running obtained for these dimensionful coupling constants   has new relevant (non-logarithmic) contributions, not predicted by dimensional regularization. %previous analysis.\\% dimensional regularization.\\

%We reexamine the adiabatic renormalization scheme by introducing and arbitrary mass scale $\mu$ in the definition of the subtraction terms. We consider the quantization of a charged scalar field and the associated pair creation process induced by an time-varying electric field in an expanding universe. We show that the modified adiabatic regularization method correctly reproduces the well-known running of the electric charge. A renormalization group analysis determines the effective form of the semiclassical Maxwell equations, which account for the backreaction effects due to the (Schwinger) pair production. We also find  significative corrections to the running of the cosmological and Newtonian  gravitational constants, as compared to that predicted by dimensional regularization. \\%These results may have implications for the cosmological constant problem.\\
%We argue that it can be relevant for the cosmological constant problem.  \\

%We investigate the adiabatic regularization and its corresponding renormalization and compute the corresponding beta functions and renormalization group for the parameters of several theories. \\


 \end{abstract}

\date{\today}
\maketitle

%The theory of quantized fields in Minkowski space is one of the most successful frameworks in physics, correctly describing the  behavior of  elementary particles in a  wide range of scales \cite{qftbook, Weinbergqft, Srednikybook}. This framework was born  with the development of   quantum electrodynamics (QED), which is regarded as the paradigm of  a perturbatively reliable quantum field theory (QFT) at low energies. %The theory predicts the existence of positrons and  naturally explains why electrons, positrons  and  photons obey the spin-statistic connection. 
\section{Introduction} Quantum field theories are intrinsically plagued with ultraviolet divergences, which need to be first isolated with the help of a regularization method and then  removed to produce finite results. This is the so-called renormalization procedure, which was  developed to prove the perturbative consistency of quantum electrodynamics (QED)  and led to very significant predictions, like the anomalous magnetic moment of the electron.  
Another remarkable prediction %, tied to the interaction between the Dirac and electromagnetic fields, 
is the effective running of the electric charge, tied to vacuum polarization phenomena.
% with the energy scale at which the theory is probed.  
The renormalization procedure typically introduces an arbitrary mass scale $\mu$, reflecting the inherent ambiguity of any scheme of renormalization \cite{Hooft2, Hooft1}. Demanding that the  bare electric charge  be independent of the renormalization scale $\mu$ leads to the effective running of the electric charge in QED: %The $\mu$ dependence of the renormalized  charge provides information on the measured effective electric charge.
$e^2(\mu^2) = e^2(\mu_0^2)/(1-\frac{e^2(\mu_0^2)}{12\pi^2} \ln \frac{\mu^2}{\mu_0^2})$. 
The choice of the reference scale $\mu_0$ at which one defines the gauge coupling will not influence physical predictions. % \cite{qftbook, Weinbergqft, Srednikybook}. 

The paradigm of QED as a quantum field theory model was generalized in two directions. On the one hand, the abelian gauge symmetry of electrodynamics was  generalized to non-abelian  symmetries to construct  a very successful model of the electroweak and strong interactions. 
%With the  aid of the dimensional regularization method, the Standard Model was proved to be a renormalizable quantum field theory \cite{Hooft1}.  
% Later on the strong interaction was also understood in terms of non-abelian gauge symmetries, having a coupling constant $g$ with a running behavior such that $g(\mu^2)$ decreases as $\mu$ increases.
On the other hand,  quantum field theory in Minkowski space was extended to curved spacetime, and the particle creation phenomena by the changing metric of an expanding universe was discovered \cite{parker66, parker68} (see also \cite{parker-toms, birrell-davies, fulling, Waldbook, Hollands}). Pairs of particles are created out of the vacuum in a non-perturbative way. Physical consistency demands that  this process should be compatible  with the covariant conservation of the stress-energy tensor of the quantized matter field. However, the formal expression for $\langle T^{\mu\nu} \rangle $ is, as  expected, afflicted by ultraviolet divergences (UV). General covariance strongly restricts the way one could construct the subtraction terms for renormalization. A physically well motivated method of regularization and renormalization, known as adiabatic regularization, was introduced in \cite{parker-fulling}. It was based on the adiabatic definition of particle states in an expanding universe, obtained by solving the  equation for the modes  in the adiabatic limit, and using a WKB-type asymptotic expansion for scalar fields \cite{parker-toms, fulling, birrell-davies}. The method has been widely used in many relevant  applications in cosmology \cite{Birrell, Bunch80, Anderson-Parker}. The adiabatic regularization method can also be adapted to deal with a quantized field in time varying electric backgrounds \cite{Cooper1},  quantized Dirac fields in an expanding universe with a Yukawa coupling \cite{rio1, Yukawa, eduardo}, and quantized fields in the presence of a time-varying  electric field in an expanding spacetime \cite{ferreiro-navarro, pla}.   These generalizations are important to account for backreaction effects in the (non-perturbative) Schwinger pair creation phenomena \cite{Schwinger}, in the lab \cite{Dunne, Ritus} and also in astrophysics and cosmology \cite{PR}, and the particle creation producing the reheating of the universe after inflation \cite{KLS}. 
%The adiabatic method shares at some level the  effective field theory strategy to separate physical scales. Some degrees of freedom are quantized while other remain considered as classical low energy backgrounds. 

However, all the  above improvements of the adiabatic method implicitly assume that the renormalization scale is  fixed at the  mass  of the quantized field.  In this work we show how an arbitrary renormalization mass scale $\mu$, playing a role somewhat similar to the conventional unit of mass $\mu$ of dimensional regularization \cite{Hooft1, Hooft2}, can be naturally incorporated into the adiabatic method. We also study the physical consequences of this novel proposal. 

To simplify the discussion we first restrict our analysis to a charged scalar field living in a spatially flat Friedmann-Lema\^itre-Robertson-Walker (FLRW) spacetime, where particle creation can also be induced  by the presence of a  background electric field. We will show how the adiabatic renormalization method predicts the running of the electric charge, exactly in the amount 
predicted by more conventional methods, such as dimensional regularization. Moreover, we will find  significant corrections to the running of the cosmological constant, $\Lambda_c$, and the Newtonian gravitational constant, $G$. %{\color{red} A renormalization group analysis permits to write down the effective  semiclassical Maxwell equations for several quantized matter fields with different masses}. 
  The running obtained for these couplings  has new relevant contributions, not predicted by dimensional regularization.  They are  linked to the intrinsic quartic and quadratic UV divergences of the stress-energy tensor.  For the extra (dimensionless)  coupling, associated with the term quadratic in the curvature, our results agree with the slow logarithmic running predicted by dimensional regularization. The method  can also be applied to quantized Dirac fields. We report here our main results. 
We use units for which $c=1=\hbar$. %Our sign conventions for the signature of the metric and the curvature tensor follow Ref. \cite{parker-toms, birrell-davies}. 

%Renormalization demanded the presence of terms quadratic in the curvature in the bare action. 

\section{Adiabatic regularization with an arbitrary mass scale $\mu$}
The main idea of the paper can be understood by considering a quantized charged scalar field living in a FLRW spacetime with metric $ds^2= dt^2 - a^2(t)d\vec x^2$, and coupled to an homogeneous electric field. 
%The complex scalar field is described by the Lagrangian density
%\bea
%\mathcal{L}=\sqrt{-g}[ (D_\mu \phi)^{\dagger}  D^\mu\phi-(m^2+ \xi R) \phi^*\phi-\frac{1}{4q^2}F_{\mu\nu}F^{\mu\nu}] \ \ \ \ \ \ .
%\eea
The complex scalar field is assumed to obey the Klein-Gordon equation $(D_\mu D^\mu + m^2 + \xi R)\phi=0$, where $D_\mu \phi = (\nabla_\mu +iA_\mu)\phi$. $\nabla_\mu$ is the covariant derivative associated with the dynamical  metric. $R$ is the Ricci scalar. For our purposes it is enough to assume that the vector potential is of  the form $A_\mu= (0, -A(t), 0, 0)$. The electric charge $q$ has been reabsorbed in the definition of $A_\mu$. Therefore, it only appears in the Maxwell Lagrangian for the pure electromagnetic field  $-\frac{\sqrt{-g}}{4q^2}F_{\mu\nu}F^{\mu\nu}$, where $F_{\mu\nu}$ is the  field strength.  %is given by $F_{\mu\nu} = \nabla_\mu A_\nu -\nabla_\nu A_\mu$. 
The quantized field can be expanded in modes of definite $3$-momentum $\vec k$\be \phi= \frac{1}{\sqrt{2(2\pi a(t))^3}} \int d^3k[A_{\vec k} e^{i\vec k\vec x}h_{\vec k}(t) + B^{\dagger}_{\vec k} e^{-i\vec k\vec x}h^*_{-\vec k}(t)] , \ee
where the time-dependence is fixed by the wave equation
\be \label{modeseq}\ddot h_{\vec k} + (\frac{(\vec k - \vec A)^2}{a^2} + m^2 + \sigma)h_{\vec k} =0 \ , \ee
and $\sigma= (6\xi -3/4) \dot a^2/a^2 +(6\xi -3/2)\ddot a/a$. 
One-particle states can only be well-defined in the adiabatic limit of slow expansion and weak electric field, where 
        \be  \label{WKB} h_{\vec k} \sim \frac{1}{\sqrt{W_{\vec k}(t)}} e^{-i\int^t W_{\vec k}(t') \mathrm{d}t'}\ . \ee
$W_{\vec k}$ is obtained by a WKB-type asymptotic  expansion in powers of $A(t)$, $\dot a(t)$; $\dot A(t)$, $\ddot a(t)$;  etc. [Note that, as explained in Refs. \cite{ferreiro-navarro, pla}, we have taken into account that $A(t)$, as well as $\dot a(t)$, are of adiabatic order 1.] Therefore, $W_{\vec k} = \omega^{(0)}_{\vec k} + \omega^{(1)}_{\vec k}+ \omega^{(2)}_{\vec k} + \cdots $ \ . The leading term, of zero adiabatic order, can be naturally taken as $\omega^{(0)}_{\vec k}\equiv  \omega= \sqrt{ \vec k^2/a^2 + m^2 }$. The next to leading term, of adiabatic order 1,  is given by  $\omega^{(1)}_{\vec k}= -Ak_x/(a^2\omega)$. 
%\bea \omega^{(2)}&=&-\frac{A^2 k_x^2}{2 a^4 \omega ^3}+\frac{A^2}{2 a^2 \omega }+\frac{3 \xi  \dot{a}^2}{a^2 \omega }-\frac{3 \dot{a}^2}{8 a^2 \omega} \nonumber \\
%&+&\frac{3 \xi  \ddot{a}}{a \omega }-\frac{3 \ddot{a}}{4 a \omega }+\frac{3 \dot{\omega }^2}{8 \omega ^3}-\frac{\ddot{\omega }}{4
 %  \omega ^2} \ . \eea
A well-defined recursion relation univocally defines the higher-order terms (for details see \cite{ferreiro-navarro}). 

A very important advantage of the above expansion
%, as first discovered in \cite{parker-fulling} in a purely gravitational setting, 
is that it allow us to identify the UV divergences emerging in the formal expectation values of non-linear operators. %, as the electric current or the stress-energy tensor.
 For instance, the formal vacuum expectation value of the  current is given by 
\bea
\langle \vec j\rangle= \int d^3k \vec j (\vec k, t) = 
\frac{1}{(2\pi)^3a^5}\int d^3k(\vec{k}+\vec A(t))|h_{\vec k}(t)|^2 \ . 
\   \ \ \ \eea
%\bea
%\langle j^0\rangle=\frac{i q}{2(2\pi)^3}\int d^3k \left(h_{\vec k}^*(t){\dot{h}}_{\vec k}(t)-{\dot{h}}_{\vec k}^*(t)h_{\vec k}(t)\right)
%\eea
The ultraviolet divergences  of the electric current $\langle j^\mu \rangle$ can be learnt from the corresponding adiabatic expansion. The mode expansion for $h_k(t)$ is  plugged in $j^\mu(\vec k, t)$ to generate an adiabatic series. One finds UV divergences up to the third adiabatic order.  The minimal number of terms in this series should be subtracted from $j^\mu(\vec k, t)$ to cancel out all UV divergences. 
 Therefore, the physical expectation values of the current are obtained by subtracting up to and including the third adiabatic order. This will give
%\bea
%\langle j^0\rangle_{ren}=0 \ , 
%\eea
%which correspond to the fact that no total charge is created by the external electric field, and
\bea
\langle \vec j\rangle_{ren}&&=\frac{1}{(2\pi)^3a^5}\int d^3k (\vec{k}+ \vec A(t))|h_{\vec k}(t)|^2\\ &&-\vec{k} (|h_{\vec k}(t)|^2)^{(0-3)}- \vec{A}(t) (|h_{\vec k}(t)|^2)^{(0-2)} \ , \ \  \  \ 
\eea
where $(|h_{\vec k}(t)|^2)^{(0-n)} =  (\omega_{\vec k}^{(0)})^{-1} + (W_{\vec k}^{-1})^{(1)}+ \cdots + (W_{\vec k}^{-1})^{(n)} $.
One also obtains $\langle j^0\rangle_{ren}=0$, 
which corresponds to the fact that no total charge is created by the external electric field. We have to stress that the above subtraction scheme is also acting as a regularization procedure. There is no need to introduce any cut-off or regularization parameter, as in any other regularization method (for instance, $\epsilon = 4-d$, in dimensional regularization). Hence the traditional adiabatic regularization name used to refer to the whole procedure \cite{parker-toms, birrell-davies, fulling}. 

A close reexamination of the above description of the adiabatic regularization procedure unravels an intrinsic ambiguity in the definition of the leading term in the adiabatic expansion of $W_{\vec k}(t)$.    The leading and fundamental term in the adiabatic expansion, $\omega^{(0)}_{\vec k}$,  can be indeed  defined in a slightly more general way  without spoiling the consistency of the overall renormalization scheme.   One can take
\be \label{wo}\omega^{(0)}_{\vec k}\equiv  \omega= \sqrt{ {\vec k}^2/a^2 + \mu^2 } \ , \ee
where $\mu$ is an arbitrary mass scale, instead of the choice $\sqrt{ \vec k^2/a^2 + m^2 }$. The higher order  adiabatic terms are univocally  recalculated as (using  (\ref{modeseq}), (\ref{WKB}), and (\ref{wo}))
\bea
\omega^{(1)}&&=-\frac{A k_x}{a^2\omega }\\
\omega^{(2)}&&=-\frac{A^2 k_x^2}{2 a^4 \omega ^3}+\frac{A^2}{2 a^2 \omega }+\frac{3 \xi  \dot{a}^2}{a^2 \omega }-\frac{3 \dot{a}^2}{8 a^2 \omega }+\frac{\alpha
   ^2}{2 \omega }\nonumber \\
   &&+\frac{3 \xi  \ddot{a}}{a \omega }-\frac{3 \ddot{a}}{4 a \omega }+\frac{3 \dot{\omega }^2}{8 \omega ^3}-\frac{\ddot{\omega }}{4
   \omega ^2}\\
\omega^{(3)}&&=-\frac{A^3 k_x^3}{2 a^6 \omega ^5}+\frac{A^3 k_x}{2 a^4 \omega ^3}+\frac{3 A \xi  \dot{a}^2 k_x}{a^4 \omega ^3}+\frac{9 A \dot{a}^2 k_x}{8 a^4
   \omega ^3}+\frac{5 A \dot{a} \dot{\omega } k_x}{2 a^3 \omega ^4}\nonumber \\ && +\frac{3 A \xi  \ddot{a} k_x}{a^3 \omega ^3}-\frac{5 A \ddot{a} k_x}{4 a^3
   \omega ^3}-\frac{\dot{a} \dot{A} k_x}{a^3 \omega ^3}+\frac{19 A \dot{\omega }^2 k_x}{8 a^2 \omega ^5}-\frac{3 A \ddot{\omega } k_x}{4 a^2 \omega
   ^4}\nonumber \\&&+\frac{\alpha ^2 A k_x}{2 a^2 \omega ^3}-\frac{5 \dot{A} \dot{\omega } k_x}{4 a^2 \omega ^4}+\frac{\ddot{A} k_x}{4 a^2 \omega ^3} \ , \eea
where $\alpha^2 \equiv m^2 - \mu^2$.  %The expansion with $\mu=m$ has been given in \cite{ferreiro-navarro}. 
The  new terms, proportional to $\alpha^2$,  serve to remove UV divergences, in accordance with the new definition of $\omega^{(0)}_{\vec k}$, while maintaining  locality and general covariance.   Note that $\alpha^2$ should be regarded as a parameter of adiabatic order 2.  The more conventional adiabatic method is recovered when the mass scale $\mu$ is fixed at the physical mass of the quantized field, i.e., $\mu=m$ and hence $\alpha=0$. The vacuum expectation values for the stress-energy tensor in Minkowski space, and in the absence of additional external fields, are then predicted to be zero. Otherwise, for a generic one $\mu$, the result is  $\langle T_{\mu\nu} \rangle_{ren} = \frac{1}{ (8\pi)^2}\left(-3m^4+4m^2\mu^2-\mu^4-4m^4\ln{\left(\frac{\mu}{m}\right)}\right)g_{\mu\nu}$. More technical details will be given elsewhere. %In an slowly expanding universe, and for an adiabatic vacuum state, the stress-energy tensor at the natural scale $\mu=m$ is of the order  $H^6/m^2$, where $H$ is the Hubble rate. \\

 %which could be reabsorbed into the cosmological constant term.}\\

%Let us remark that  the traditional adiabatic method (with $\mu=m$) has been proved \cite{Anderson-Parker, rio-navarro} to be equivalent to the DeWitt-Schwinger point-splitting method. In momentum space the latter is based on an adiabatic  expansion of the two-point function. For a scalar field one has \cite{Bunch-Parker} $ \bar G(p) = (-k^2 + m^2)^{-1} + (1/6 -\xi) R(-k^2 + m^2)^{-2} + ... \ . $ Our previous discussion suggests, as a by-product, that one can also introduce an arbitrary mass scale to modify the above expansion, mimicking the introduction of the mass scale $\mu$ into the adiabatic regularization method. 
%by a shifting of the expansion similar to the %one proposed here for the adiabatic method 
%\be \bar G(p) = \frac{1}{-k^2 + \mu^2} + \frac{(\mu^2 -m^2) + (1/6 -\xi) R}{(-k^2 + \mu^2)^2} + ... \ee
%We expect that the this  generalized expansion be equivalent to the above generalized  adiabatic method

\section{Renormalization of the electric current and the running of the electric charge}
The introduction of the mass scale $\mu$ leads to an inherent ambiguity in the adiabatic renormalization scheme, as also happens in dimensional regularization. It is natural to compare the renormalized  current at two different scales: $\langle j^{\beta}\rangle_{\rm ren}(\mu) -  \langle j^{\beta} \rangle_{\rm ren}(\mu_0) = \langle j^{\beta}\rangle^{(0-3)}(\mu_0)-\langle j^{\beta}\rangle^{(0-3)}(\mu)$. 
By using the above adiabatic expansion we find (we rewrite the result in  covariant terms)

\bea
&\langle j^{\beta}\rangle_{\rm ren}(\mu) -  \langle j^{\beta} \rangle_{\rm ren}(\mu_0) =-2\delta_q \nabla_{\alpha}F^{\alpha\beta}\label{currentmmu2},
\eea
with $\delta_q =\frac{1}{ 3 (4\pi)^2} \ln{(\frac{\mu}{\mu_0})}$. % and $\nabla_{\mu}\sqrt{-g}F^{\mu1}=\dot{a}\dot{A}+a\ddot{A}$. By using \eqref{Maxwell} and \eqref{currentmmu} we obtain
The semiclassical Maxwell equations should take the form, irrespective of the value of renormalization parameter $\mu$,
\bea
\frac{1}{q^2(\mu)}\nabla_{\alpha}F^{\alpha\beta}=\langle j^{\beta}\rangle_{ren}(\mu) \ . \label{Maxwell1}
\eea
Therefore, we must also have
\bea
\frac{1}{q^2(\mu_0)}\nabla_{\alpha}F^{\alpha\beta}=\langle j^{\beta}\rangle_{ren}(\mu_0) \ . \label{Maxwell2}
\eea
Demanding now physical equivalence between  (\ref{Maxwell1}) and (\ref{Maxwell2}), and using (\ref{currentmmu2}), one obtains the  running of the electric charge
\be \label{rq}
\frac{1}{q^2(\mu)}-\frac{1}{q^2(\mu_0)}= -\frac{1}{ 48\pi^2} \ln\frac{\mu^2}{\mu_0^2}\  , \ee
in full agreement with the result obtained within perturbative scalar QED in Minkowski space (using, for instance, dimensional regularization and the modified minimal subtraction scheme  \cite{Srednikybook}). 
Note that, for getting the above result, there has been no need to assume a generic form for the electromagnetic background. It has been enough to use a background potential of the form $A_\mu= (0, -A(t), 0, 0)$. We also remark that (\ref{rq}) has been obtained  without using any perturbative expansion in the coupling constant $q$.
 There are no corrections to the running of the gauge coupling coming from classical gravity.

\section {Renormalization of the stress-energy tensor and the running of the gravitational couplings}
The formal vacuum expectation value of the stress-energy tensor $\langle T_{\mu\nu} \rangle$ possesses UV divergences up to adiabatic order 4, inclusive. Therefore, the physical expectation values 
 $\langle T_{\mu\nu} \rangle_{ren}$ are obtained by subtracting up to and including the fourth adiabatic order. Comparing now  $\langle T_{\mu\nu} \rangle_{ren}(\mu)$ and  $\langle T_{\mu\nu} \rangle_{ren}(\mu_0)$ we find (for simplicity we take here $\mu_0=m$)
 
 \bea
\langle T_{\alpha\beta}\rangle_{\rm ren}(\mu) -  &&\langle T_{\alpha\beta} \rangle_{\rm ren}(m)=\delta_{\Lambda} g_{\mu\nu}+\delta_{G} G_{\mu\nu}+\delta_{\alpha^1} H^{(1)}_{\mu\nu} \nonumber \\
&&+2\delta_q \left(\frac14 g_{\alpha\beta}F_{\rho\sigma}F^{\rho\sigma}-F_{\alpha}^{ \ \rho}F_{\beta\rho}\right) \label{eqummu2}
\eea
with
\bea 
&&\delta_{G} =\frac{-1}{ (8\pi)^2} \frac43\left(1-6\xi\right)(m^2-\mu^2-2m^2\ln {\frac{m}{\mu}})\\
&&\delta_{\alpha^1} =\frac{-1}{ (8\pi)^2}\frac29\left(1-6\xi\right)^2 \ln {\frac{m}{\mu}} \\
&&\delta_{\Lambda}=\frac{1}{ (8\pi)^2}(-3m^4+4m^2\mu^2-\mu^4-4m^4\ln {\frac{\mu}{m}})\\
&&\delta_q =\frac{1}{ 3 (4\pi)^2} \ln {\frac{\mu}{m}}
\ . \eea
$ H^{(1)}_{\mu\nu}$ is the conserved curvature tensor obtained by functionally differentiating the quadratic curvature lagrangian $R^2$ with respect to the metric. The extra term $\delta_{\alpha^1} H^{(1)}_{\mu\nu}$  implies the existence of a modification of general relativity due to quantum effects, as first pointed out in Ref. \cite{Utiyama} for asymptotically flat spacetimes.  Here there is no need to introduce the additional conserved tensor,
$ H^{(2)}_{\mu\nu}$, 
coming from the lagrangian $R_{\mu\nu}R^{\mu\nu}$. %$C_{\mu\nu\rho\sigma}C^{\mu\nu\rho\sigma}$, where $C$ is the Weyl tensor. 
This is because,  in a FLRW  spacetime, $H^{(1)}_{\mu\nu}$ and $H^{(2)}_{\mu\nu}$ are not independent.  As long as we treat the gravitational field as a classical background, no terms of higher order in the curvature are required.

At this point we should remark that  expression (\ref{eqummu2}) is compatible with the ambiguities in the quantization of the stress-energy tensor found in the algebraic approach to QFT in curved spacetime \cite{Waldbook, Hollands, Wald-Hollands01}. To be more precise, any two local and covariant procedures of renormalization of the stress-energy tensor should differ at most in a linear combination of conserved local terms: $a_1 g_{\mu\nu}+ a_2 G_{\mu\nu} + \alpha_1 H_{\mu\nu}^{(1)} + \alpha_2 H_{\mu\nu}^{(2)}$. %The ambiguities in $a_1$ and $a_2$ can be absorbed into the conslo to the se ambiguities are expected to be solved by observations or by a quantum theory of gravity \cite{Waldbook}.
In a FLRW spacetime, $H_{\mu\nu}^{(2)}$ is proportional to  $H_{\mu\nu}^{(1)}$, hence $\alpha_2$ can be reabsorbed into $\alpha_1$. Moreover, since we have an additional external field (the electromagnetic background), the ambiguity should also include the  electromagnetic stress-energy tensor. %\left(\frac14 g_{\alpha\beta}F_{\rho\sigma}F^{\rho\sigma}-F_{\alpha}^{ \ \rho}F_{\beta\rho}\right)$.
 Therefore, given two prescriptions to renormalize the stress-energy tensor,  denoted by $\langle T_{\alpha\beta}\rangle_{\rm ren}$  and  $\langle \tilde T_{\alpha\beta}\rangle_{\rm ren}$, the difference for the expected stress-energy tensor is parametrized by the following linear combination
 \bea
\langle T_{\alpha\beta}\rangle_{\rm ren} -  \langle \tilde T_{\alpha\beta} \rangle_{\rm ren}= &&a_1g_{\alpha\beta}+a_2G_{\alpha\beta}+\alpha_1 H^{(1)}_{\alpha\beta} 
\nonumber \\ &&+ a_3 \left(\frac14 F_{\sigma\rho}F^{\sigma\rho}g_{\alpha\beta}-F_{\alpha}^{ \ \rho}F_{\beta\rho}\right)  \ . \ \  \ \ \ \label{eqummu2bis}
\eea 
The constant parameters $a_1, a_2, a_3$ and $\alpha_1$ are not constrained within the axiomatic approach. 

We can identify now $\langle \tilde T_{\alpha\beta} \rangle_{\rm ren}$ with the standard adiabatic prescription to renormalize the stress-energy tensor  $\langle \tilde T_{\alpha\beta} \rangle_{\rm ren} \equiv \langle T_{\alpha\beta} \rangle_{\rm ren}(m)$, and $\langle T_{\alpha\beta}\rangle_{\rm ren}$ with our modified adiabatic prescription (parametrized by the mass scale $\mu$): $\langle T_{\alpha\beta}\rangle_{\rm ren} \equiv \langle T_{\alpha\beta}\rangle_{\rm ren} (\mu)$. Therefore, the constant and finite parameters $a_1, a_2, a_3$ and $\alpha_1$ naturally acquire a dependence on the scale $\mu$. This dependence can be obtained by direct computation and the result is given by the above expressions for $\delta_{\Lambda}$, $\delta_G$, $\delta_{q}$ and $\delta_{\alpha_1}$, respectively.  Furthermore, as we will see now, this implies a natural running for the  gravitational coupling constants.

The semiclassical Maxwell-Einstein equations are given by (\ref{Maxwell1}) together with 
\be \langle T_{\alpha\beta}\rangle_{ren}(\mu)+T^{EM}_{\alpha\beta}(\mu)=\frac{-G_{\alpha \beta}}{8\pi G(\mu)}-\Lambda(\mu) g_{\alpha \beta} - \alpha^1(\mu)  H^{(1)}_{\alpha\beta}\ , \ee
with $T_{\alpha\beta}^{EM}=\frac{1}{q^2(\mu)}\left(\frac14 F_{\sigma\rho}F^{\sigma\rho}g_{\alpha\beta}-F_{\alpha}^{ \ \rho}F_{\beta\rho}\right)$. The coupling $\Lambda$ is related to the cosmological constant $\Lambda_c$ by the relation  $\Lambda =\Lambda_c/(8\pi G)$. Enforcing  that the above equations be independent of the scale $\mu$, we obtain, using the above results for $\delta_{G}$ and $\delta_{\Lambda}$, the running of the Newton gravitational constant $G$ and $\Lambda$. The running of $q$ can also be obtained, and coincides with the result (\ref{rq}), derived directly from the renormalization of the electric current. 
We find 
%\be \Lambda(\mu) - \Lambda(m) = \frac{1}{ (8\pi)^2}\left(-4m^2\mu^2+3m^4+\mu^4+2m^4\log{\frac{\mu^2}{m^2}}\right)\ee
 \be \label{r1}\Lambda(\mu) - \Lambda(\mu_0) = \frac{1}{ (8\pi)^2}(\mu^4 - \mu_0^4-4m^2(\mu^2- \mu_0^2)+2m^4\ln{\frac{\mu^2}{\mu_0^2}})\ee
\be \label{r2}\frac{1}{16\pi G(\mu)} - \frac{1}{16\pi G(\mu_0)}= \frac{(\frac{1}{6}-\xi)}{ (4\pi)^2} (\mu_0^2-\mu^2+m^2\ln{\frac{\mu^2}{\mu_0^2}}) \ee
\be \label{r3} \alpha^1(\mu) - \alpha^1(\mu_0)=    \frac{1}{ (4\pi)^2} (\frac{1}{6}-\xi)^2\ln{\frac{\mu^2}{\mu_0^2}}   \ . \ee
%{\color{blue}\bea&&  \Lambda(\mu) - \Lambda(\mu_0) = \nonumber\\ && \frac{1}{ (8\pi)^2}\left(\mu^4 - \mu_0^4-4m^2(\mu^2- \mu_0^2)+2m^4\log{\frac{\mu^2}{\mu_0^2}}\right) \\ && \frac{1}{16\pi G(\mu)} - \frac{1}{16\pi G(\mu_0)}=\nonumber \\&& \frac{1}{ (4\pi)^2}\left(\frac{1}{6}-\xi\right) \left(\mu_0^2-\mu^2+m^2\log{\frac{\mu^2}{\mu_0^2}}\right) \\&& \alpha^1(\mu) - \alpha^2(\mu_0)=    \frac{-1}{ (4\pi)^2} (\frac{1}{6}-\xi)^2\log{\frac{\mu^2}{\mu_0^2}}   \ . \eea}
We observe that the logarithmic terms in the above expressions  coincide exactly with the running predicted  by dimensional regularization (see  \cite{parker-toms}, section 6.7). However,  we find a corrected behavior for the running of $G$ and $\Lambda$, derived from (\ref{eqummu2}). The polynomial terms in the mass scales are associated to the   quadratic and quartic UV divergences of the stress-energy tensor of the matter field.  These contributions to the running of dimensionful constants  are not  captured, as expected, by a mass-independent subtraction scheme, like dimensional regularization with minimal subtraction. For the dimensionless coupling constants (electric charge $q$ or $\alpha_1$) the adiabatic method and dimensional regularization lead to the same running. 

It is interesting to relate the non-logarithmic  contributions to the running of the Newton constant with the results of the asymptotic safety approach to quantum gravity \cite{review} (see also \cite{polyakov}). The running obtained within quantum gravity is often encapsulated by the expression
\be G(\mu) = \frac{G(\mu_0)}{1 + a G(\mu_0) (\mu^2 - \mu^2_0)} \ . \ee
This result coincides with the one predicted by (\ref{r2}) for negligible mass, with $a=(\xi -1/6)/\pi$.

%{\color{red}  Following the argument that leads to \eqref{Maxwell3} one can also determine the form of the semiclassical Einstein equations. For illustrative purposes, if one consider the simplest case of two scalar fields with masses $m_1$ and $m_2$,  the effective semiclassical Einstein equations take the form \bea &&[\frac{1}{8\pi G(m_1)} - \frac{4}{3(8\pi)^2}(1-6\xi_2)(m_2^2 - m_1^2 -2m_2^2 \log \frac{m_2}{m_1}]G_{\alpha\beta} \nonumber \\ &&[\Lambda(m_1)- \frac{1}{(8\pi)^2}(-3m_2^4 + 4m_2^2m_1^2 -m_1^4 -4 m_2^4 \log \frac{m_1}{m_2})]g_{\alpha\beta} \nonumber \\ &&= -\langle T_{\alpha\beta}\rangle_{ren} (m_1) - \langle T_{\alpha\beta}\rangle_{ren} (m_2) \eea}

%{\color{red}The renormalization group transformations in curved spacetime is related to a scaling of the metric \cite{Nelson-Panangaden, parker-tomsPRL, Hollands-WaldCMP}, and hence of curvature invariants.Since $G^{-1}$ is very large, the running  does not significantly modify the current value of Newton's constant. Only at curvatures of order of  the Planck scale one could expect a  significant change.  On the contrary, the smallness value of the observed effective cosmological constant at low curvature could be largely altered, due to the new non-logarithmic contributions, at large curvatures. \\}
%In contrast with the semiclassical Maxwell equations in (\ref{Maxwell3}), the details of the left-hand-side of the equations are not so useful. Due to the lack of empirical data (we do not have precise experiments to independently  

\section{Running couplings, Dirac fields, and backreaction equations} The details of the  scaling behavior depend on the type of field. For a Dirac field $\psi$ the computations are more involved, since in this case the adiabatic expansion of the fermionic modes requires a generalization of the WKB-type ansatz \cite{rio1, Yukawa, eduardo, ferreiro-navarro}.  Our calculations yield  
%(we give here only the results for the relevant gravitational couplings)
\bea \label{Lambdar}&& \Lambda(\mu) - \Lambda(\mu_0)=   \frac{-1}{ 96 \pi^2} [ 48m^3(\mu_0-\mu)+36m^2(\mu^2-\mu_0^2) \nonumber \\&&-16m(\mu^3-\mu_0^3)+3(\mu^4-\mu_o^4)+ 6m^4\ln{\frac{\mu^2}{\mu_0^2}}]\eea
\bea \label{Gr}&& \frac{1}{16\pi G(\mu)} - \frac{1}{16\pi G(\mu_0)}=\nonumber \\&& \frac{1}{ 32 \pi^2} [4 m(\mu-\mu_0)-(\mu^2-\mu_0^2)-m^2\ln{\frac{\mu^2}{\mu_0^2}}] \ ,  \eea
while for the electric charge we have the standard result for a Dirac fermion $ q^{-2}(\mu)-q^{-2}(\mu_0)= -\frac{1}{ 12\pi^2} \ln\frac{\mu^2}{\mu_0^2}$. 

%For a set of different species of fields one can also write the semiclassical Einstein equations in a consistent way with the renormalization group argument, as we have already done for the semiclassical Maxwell equations. 
 % {\color{blue} We also mention that we have checked that the improved adiabatic method exactly reproduces the  results for the running of other typical coupling constants of QFT in Minkowski space, like the Yukawa coupling $g_Y$ measuring the strength of the interaction between a Dirac field and an external scalar field $g_Y \bar\psi \psi \Phi$. %To check this we have updated, introducing a mass scale $\mu$, the results of \cite{Yukawa}. The details will be given elsewhere.}
 
%A major advantage of our renormalization group  arguments emerges when one.  
Let us now  consider $N$ charged  fields $\psi_i$ with masses $m_i$. For simplicity we take all fields with the same electric charge. In this case the semiclassical Maxwell equations should read $q^{-2}(\mu)\nabla_{\alpha}F^{\alpha\beta}=\sum_{i=1}^N \langle j_i^{\beta}\rangle_{ren}(\mu)$. As above, we demand these equations be independent of $\mu$. %Since, as argued above, these equations should be independent of the renormalization scale $\mu$, 
Hence, the running of the coupling is now found $ q^{-2}(\mu)-q^{-2}(\mu_0)= -\frac{N}{ 12\pi^2} \log\frac{\mu^2}{\mu_0^2}$. The  final form of the  effective semiclassical equations turns out to be 
%\bea
%\frac{1}{N} (\sum_{j=1}^N\frac{1}{q^2(m_j)} )%+ \frac{1}{6\pi^2} \ln \frac{m_j^{N-1}}{m_1 m_2 \hat {m_j} m_N})
%\nabla_{\alpha}F^{\alpha\beta} =
%\sum_{i=1}^N\langle j_i^{\beta}\rangle_{ren}(m_i) \ , \label{Maxwell3}
%\eea
\be(\frac{1}{q^2(m_j)}+ \frac{1}{6\pi^2} \ln \frac{m_j^{N-1}}{m_1 m_2 \hat {m_j} m_N})\nabla_{\alpha}F^{\alpha\beta} =
\sum_i\langle j_i^{\beta}\rangle_{ren}(m_i) \ ,  \ \ \ \ \ \label{Maxwell3}
\ee
where the  currents $\langle j_i^{\beta}\rangle_{ren}$ have been renormalized  at their respective natural mass scales $m_i$. This is the origin of the second term inside the parenthesis on the left-hand-side in (\ref{Maxwell3}). It is easy to check that the above equations are  indeed independent of the reference mass $m_j$. These equations are univocally fixed once we give the masses and the experimental value of the charge at the reference scale, $q^2(m_j)$ ($m_j$ can the taken as the smallest value of the masses).
%It is easy to check that the above equations are  indeed independent of the reference mass $m_j$ used on the left-hand-side.} 

  Following the argument that leads to \eqref{Maxwell3} one can also determine the form of the semiclassical Einstein equations for $N$ species of fields with different masses. The effective semiclassical Einstein equations take the form
%\be  \frac{1}{N} [\sum_{i=1}^N \frac{1}{8\pi G(m_j)}  G_{\alpha\beta} + \sum_{i=1}^N \Lambda(m_j)g_{\alpha\beta} ] = -\sum_{i=1}^N \langle T_{i  \alpha\beta}\rangle_{ren} (m_i)  \ee 
\bea &&[\frac{1}{8\pi G(m_j)} - f_j(m_k)]G_{\alpha\beta} + 
 [\Lambda(m_j)- g_j(m_k)]g_{\alpha\beta} =\nonumber \\
 && -\sum_{i=1}^N \langle T_{\alpha\beta}\rangle_{ren} (m_i)  \eea
where $f_j(m_k)$ and $g_j(m_k)$ are functions of the masses that can be derived, for Dirac fields, from (\ref{Gr}) and (\ref{Lambdar}). The equations are also independent of the reference scale $m_j$ used. 
%The running of the coupling constants are given by (\ref{Lambdar}) and  (\ref{Gr}), where $m$ is replaced by $\sum_i m_i$. 
Since the observed value of $(8\pi G)^{-1}$ is very large, the running  does not significantly modify the  value of $(8\pi G)^{-1}(m_j)$, unless very large masses are added to the spectrum. %Only at curvatures of order of  the Planck scale one could expect a  significant change.  
 On the contrary, the  value of the observed effective cosmological constant depends, in an involved way, on the pattern of masses. It also depends on the value of $\Lambda(m_j)$, where $m_j$ is a  reference mass. % which can be taken as the smallest value. 
  In a more realistic scenario one should include the whole field content of the Standard Model.
 
 % including the vacuum contribution of the electroweak symmetry breaking and perhaps a potential non-zero vacuum contribution from  non-perturbative physics.  
 %This is of course an extremely difficult and open problem.  However, the  functions $g_j$ may conspire to produce a very tiny value  out of the mass spectrum of the Standard Model. it may be possible \\
%In contrast with the semiclassical Maxwell equations in (\ref{Maxwell3}), the details of the left-hand-side of the equations are not so useful. 
In the case of electrodynamics we have empirical data for  $\alpha (m_e)$ (using, for instance, the Josephson effect or  Thomson scattering). Hence we can predict the effective value for $\alpha$ in the semiclassical Maxwell equations for very strong electric fields, capable to create different species of charged particles. %, describing, for instance, the physics of the Schwinger pair creation phenomena. 
Unfortunately, and in sharp contrast to the case of electrodynamics, we do not have direct  data for $\Lambda(m_j)$  and this makes  it extremely difficult to produce specific observable  predictions for the effective cosmological constant.

\section{Conclusions and final comments}  In this work we have generalized the adiabatic regularization method to accommodate, in a natural way, an arbitrary renormalization mass scale $\mu$. The overall renormalization method maintains  covariance and locality and also retains its practical advantage  for numerical calculations. We have shown how the renormalization group flow for coupling constants emerges within the adiabatic method. Concerning the scaling behavior of the electric charge, we have  reproduced the well-known fact (\ref{rq}) in a new way, and without invoking  any perturbative argument.  %This can be regarded as a  further non-trivial test of the  consistency and robustness  of the adiabatic framework.
 
 One of the main advantages of the adiabatic approach is that  it  determines the effective  semiclassical Maxwell equations for  several quantized matter fields with different masses, as stated in (\ref{Maxwell3}). For instance, in a scenario where the electric field is strong enough to  create electron-positron (and 
muon-antimuon) pairs, the semiclassical (backreaction) Maxwell equations are
$\alpha^{-1} \nabla_\sigma F^{\sigma\beta } = \langle j_e^{\beta}\rangle_{ren}(m_e) +  \langle j_{\mu}^{\beta}\rangle_{ren}(m_{\mu})$. 
The inverse of the effective fine structure constant $\alpha\equiv e^2/4\pi$ in the above equations %+  \langle j_{\tau}^{\beta}\rangle_{ren}(m_{\tau})
 is found to be, according to (\ref{Maxwell3}),  $\alpha^{-1} = \alpha^{-1}(m_e) + \frac{2}{3\pi} \ln (m_e/m_{\mu}) \approx 135.9$,  instead of the conventional value, $\alpha^{-1}(m_e) \approx 137.0$\ . %, if one ignores the existence of the heavy leptons. 
 %{\color{green} Razon: $137.036 - \frac{2}{3\pi}log(\frac{0.511^2 MeV^2}{105.66 MeV 1776.86 MeV}=134.174$}
  
 %For instance, in the case of fermions and for the set of three charged leptons (electron, muon, tau), the inverse of the effective fine structure constant in the semiclassical Maxwell equations for the Schwinger pair creation effect $\alpha^{-1} \partial_\sigma F^{\sigma\beta } = \langle j_e^{\beta}\rangle_{ren}(m_e) +  \langle j_{\mu}^{\beta}\rangle_{ren}(m_{\mu}) +  \langle j_{\tau}^{\beta}\rangle_{ren}(m_{\tau}) $, is found to be $\alpha^{-1} \approx 135.6 \ , $ instead of the conventional value, $137$, if one ignores the existence of the heavy leptons. 

Furthermore, the adiabatic method allows us to compute, in a very direct way, the running of the gravitational constants. 
%Note that the method was originally  constructed to deal with UV divergences in a gravitational setting and it is based on the adiabatic definition of particles in a time-varying gravitational field.  
 Here the adiabatic method gives results that differ from those found with dimensional regularization, although it agrees  
 exactly as far as the logarithmic pieces of the running constants are concerned \cite{parker-toms}. We have found additional contributions which are quadratic and quartic in the mass scale $\mu$. %physical mass of the fields. 
 For instance, for very large $\mu$ the running found for  $G(\mu)$ fits well with the predictions of some quantum gravity approaches \cite{review, polyakov}.

 The reason of the discrepancy with dimensional regularization could be traced back to the fact that dimensional regularization is not sensitive to the non-logarithmic  UV divergences of  the stress-energy tensor. On the contrary, the adiabatic subtractions are constructed to fit all types of UV divergences, while maintaining general covariance and locality. We also note that this is not a solely property of the adiabatic regularization method. Let us remark that  the conventional adiabatic renormalization method (with $\mu=m$) has been proved \cite{Anderson-Parker, rio} to be equivalent to the DeWitt-Schwinger point-splitting method. In momentum space the latter is based on the Bunch-Parker adiabatic  expansion of the two-point function. For a scalar field one has \cite{Bunch-Parker, parker-toms} 
 \be  \bar G(k) = \frac{1}{(-k^2 + m^2)} + \frac{(1/6 -\xi) R}{(-k^2 + m^2)^2} + ... \ . \ee Our  discussion suggests, as a by-product, that one can also introduce an arbitrary mass scale to modify the above expansion, mimicking the introduction of the mass scale $\mu$ into the adiabatic regularization method. Therefore, one has the generalized expansion
  \be  \label{BPmu} \bar G_{\mu}(k) = \frac{1}{(-k^2 + \mu^2)} + \frac{(\mu^2-m^2 +(1/6 -\xi) R)}{(-k^2 + \mu^2)^2} + ... \ . \ee 
 We have checked  that this new $\mu$-dependent  expansion is equivalent to the adiabatic expansion in the modes introduced in this work. The predictions of the expansion (\ref{BPmu})  for the running of the couplings should be then equivalent to those presented in this work. 
%by a shifting of the expansion similar to the %one proposed here for the adiabatic method 
%\be \bar G(p) = \frac{1}{-k^2 + \mu^2} + \frac{(\mu^2 -m^2) + (1/6 -\xi) R}{(-k^2 + \mu^2)^2} + ... \ee
%We expect that the this  generalized expansion be equivalent to the above generalized  adiabatic method

 Finally, we note that there is no universal agreement in the literature on the actual renormalization flow  for the gravitational coupling constants, nor on the gravitational corrections to the running of the gauge couplings \cite{Robinson-Wilczek, Toms, Ander-Donoghue}, and the physical interpretation of the running \cite{Ander-Donoghue, Sola}. We may expect that, future empirical inputs at high curvatures or energies, either from very early cosmology or  localized  strong gravity events (as such detected by the LIGO-Virgo collaboration), may  test the  theoretical running of the gravitational couplings.

{\it Acknowledgments.--} We thank I. Agullo, J. F. Barbero G., A. Del Rio, F. Torrenti and  E.J.S. Villase\~{n}or for useful comments on the manuscript. %J. N.-S. has benefited from many  discussions with L. Parker. 
A. F. thanks S. Antusch for his hospitality at the Physics Department of the University of Basel while  part of this work was carried out.  This work has been supported by research Grants % Grants. No.\ FIS2014-57387-C3-1-P ; 
No.\  FIS2017-84440-C2-1-P; No. \ FIS2017-91161-EXP; No. \ SEJI/2017/042 (Generalitat Valenciana); COST action CA15117 (CANTATA), supported by COST (European Cooperation in Science and Technology); No. \  SEV-2014-0398. A. F. is supported by the Severo Ochoa Ph.D. fellowship, Grant No. SEV-2014-0398-16-1, and the European Social Fund. Most of the computations have been done with the help of {\it Mathematica.}

\end{document}